# Athermal and tunable echelle grating wavelength demultiplexers using a Mach-Zehnder interferometer launch structure

DANIELE MELATI,[1,2] DAN-XIA XU,[1] ROSS CHERITON,[1] SHURUI WANG,[1] MARTIN VACHON,[1] JENS H. SCHMID,[1] PAVEL CHEBEN,[1] SIEGFRIED JANZ,[1,*]

[1]*Advanced Electronics and Photonics Research Centre, National Research Council Canada, Building M-50, 1200 Montreal Rd. Ottawa, ON, Canada, K1A 0R6*
[2] *Centre de Nanosciences et de Nanotechnologies, CNRS, Université Paris-Saclay, 91120 Palaiseau, France*
*\*Siegfried.janz@nrc-cnrc.gc.ca*

**Abstract:** We present a comparative experimental study of three silicon photonic echelle grating demultiplexers that are integrated with a Mach-Zehnder interferometer (MZI) launch structure. By appropriate choice of the MZI configuration, the temperature induced shift of the demultiplexer channel wavelengths can be suppressed (athermal) or enhanced (super-thermal), or be controlled by an on-chip micro-heater. The latter two configurations allow the channel wavelengths to be actively tuned using lower power than possible by temperature tuning a conventional echelle demultiplexer. In the athermal configuration, the measured channel spectral shift is reduced to less than 10 pm/°C, compared to the 83 pm/°C shift for an unmodified echelle device. In super-thermal operation an enhanced channel temperature tuning rate of 170 pm/°C is achieved. Finally, by modulating the MZI phase with an on-chip heater, the demultiplexer channels can be actively tuned to correct for ambient temperature fluctuations up to 20 °C, using a drive current of less than 20 mA.

## 1. Introduction

The temperature-dependent behavior of integrated photonic devices is a critical aspect of device performance and system design. This is particularly important for silicon-on-insulator (SOI), III-V semiconductor, and polymer waveguides which have very large thermo-optic coefficients [1-3]. When these materials are used in wavelength selective demultiplexers and add-drop filters deployed in wavelength division multiplexing (WDM) systems, the channel wavelengths may change by up to 100 pm/°C, resulting in channel insertion loss fluctuations, cross-talk and signal distortion [2,3]. A 10 °C temperature drift can shift the channel passbands by more than a full WDM wavelength grid spacing in a dense WDM network. Therefore active temperature control is usually necessary to maintain a photonic device at the correct operating point in a fluctuating environment. Active temperature control requires significant power to operate and dissipates even more heat to the immediate environment. In order to limit this power consumption and heat dissipation, several temperature independent (i.e. athermal devices) have been proposed [4-12]. In athermal devices the channel wavelengths ideally remain fixed as temperature changes so that active stabilization is unnecessary. On the other hand, an athermal device must operate exactly to specification as fabricated, since any intrinsic channel wavelength offsets cannot be corrected by temperature tuning. Fabrication to the required dimensional tolerances can be challenging, particularly for high index contrast silicon devices. An alternative strategy is to design actively tunable devices with power efficient tuning mechanisms. The goal in this approach is to reduce the power consumption needed to maintain the device at the correct temperature, while retaining the ability to tune a device that is slightly off specification back to the correct operating point. The fabrication tolerances may therefore be much less demanding than for an equivalent athermal device.

In this paper we demonstrate three related echelle grating demultiplexer configurations that can provide athermal performance or alternatively, enhanced tunability with low power consumption, all using a co-integrated Mach-Zehnder interferometer (MZI) structure [13]. While we focus on echelle grating devices (14-17), the same approach is also applicable with little modification to arrayed waveguide grating (AWG) demultiplexers [17-21]. The design of a silicon-on-insulator (SOI) athermal echelle grating demultiplexer with a temperature-synchronized MZI launch structure was first proposed and analyzed by Melati [13]. The MZI is used to compensate for the temperature induced diffraction angle shift of the echelle grating, thereby suppressing the temperature dependence of the transmission spectrum. Here we provide the first experimental demonstration of this athermal design [13]. This paper also extends the theory in Melati [13] to include an analysis of the role of group index dispersion on the performance and operating range of the MZI-echelle devices, and also the crucial relationship between index dispersion and fabrication tolerances. Finally, this paper extends the previous work by adapting the MZI launch structure to two new thermally tunable MZI-echelle designs that use significantly less tuning power than would be required to tune the channel wavelengths of a conventional echelle grating. The super-thermal configuration is identical to the athermal case, but with the MZI orientation reflected with respect to the echelle. In this configuration the MZI adds to the intrinsic temperature shift of the echelle diffraction angle, rather than cancelling the shift, thereby increasing the echelle temperature response. In the third configuration a local heater is added to one arm the MZI launch structure. This echelle can be actively tuned independently of the underlying substrate temperature. Since the micro-heater heats only a very small volume of silicon, the device can be tuned using far less power than would be needed to adjust the temperature of the entire chip and underlying mount. Individual devices on the same chip can also be tuned independently, to accommodate thermal gradients and intra-chip fabrication variations.

## 2. Theory and design

All the experimental devices used in this work are based on the silicon-on-insulator (SOI) echelle grating design shown in Fig. 1. The design is similar to a previously reported four channel echelle O-band device [15], which has here been modified for operation in the telecommunications C-band between $\lambda$ = 1530 nm and 1565 nm. The echelle is designed to operate using TE polarized waveguide modes. The incident and diffracted beam angles $\theta$ and $\varphi_k$ relative to the grating normal satisfy the grating equation [16],

$$\Lambda(sin\theta + sin\varphi_k) = m \left(\frac{\lambda_k}{N_{eff}}\right), \qquad (1)$$

where $m$ is the order of the grating, $\Lambda$ is the grating pitch, $\lambda_k$ is the wavelength of the $k^{th}$ output channel, and $N_{eff}$ is the corresponding effective index of the slab waveguide at $\lambda_k$. $N_{eff}$ is temperature dependent due to the strong thermo-optic effect in silicon ($dn_{Si}/dT \sim 1.8 \times 10^{-4}$ at $\lambda$=1550 nm [1]) so for the basic echelle demultiplexer of Fig. 1 the output channel wavelengths will increase with temperature. For the commonly used 220 nm thick Si waveguide core layer the thermo-optic shift is approximately 80 pm/K near $\lambda$ = 1550 nm for TE polarized light.

For the devices fabricated in this work, all the Si waveguides have a 220-nm thick Si core layer with silicon dioxide upper and lower claddings. The nominal single mode waveguide width throughout the chip is 500 nm, except where noted otherwise. The demultiplexer is designed to have four wavelength channels at an 800 GHz frequency spacing (equivalent to 6.4 nm in wavelength) across the C-band. The grating layout is a Rowland circle configuration with a grating radius of 250 μm, a diffraction order of 20, and a mounting angle of 45°. The overall echelle area is 440 μm by 210 μm, not including the MZI launch input structure. As in the previously reported O-band demultiplexer [15], the grating facets employ Bragg grating reflectors to ensure a high facet reflectivity within the C-band. These reflectors consist of a 3.3 μm long array of trenches fully etched through the silicon layer, with a grating period of 327 nm and a trench width of 115 nm. The basic echelle demultiplexer in Fig. 1 has a single input waveguide that tapers out to a 2 μm wide aperture where it joins the slab waveguide free propagation region (FPR) of the echelle grating. The four output waveguides also have a 2 μm aperture at the focal line where they connect to the slab section. These widths determine the waveguide mode profile launched into the FPR and the output waveguide mode overlap with the diffracted beam profiles, and hence determine the channel passband shape and width. The resulting 3-dB bandwidth of each channel is 250 GHz (2 nm in wavelength).

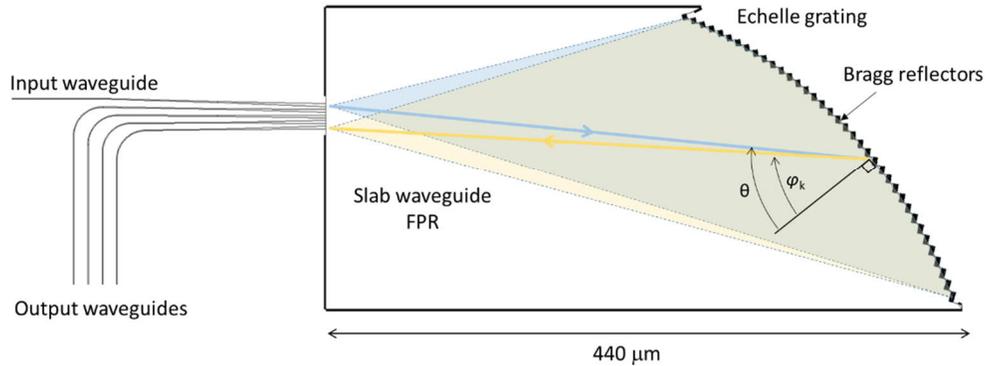

Fig. 1. Layout of the basic four channel echelle grating demultiplexer with a single input waveguide and four output channel waveguides. Light is launched into the slab waveguide free propagation region (FPR) from the input waveguide towards the echelle grating, which diffracts and focuses the light backwards onto the output waveguide array.

In all of the athermal, super-thermal and tunable MZI-echelle devices, the single input waveguide in Fig. 1 is replaced by the coupled waveguide pair that forms the output 3 dB directional coupler (DC2) of the MZI, as described by Melati [13]. This MZI structure is shown schematically in Fig. 2, and details of the fabricated on-chip layout are shown in Fig. 3. The MZI uses a 3-dB directional coupler power splitter (DC1) at the input side to couple light into two single mode Si waveguide arms. The light in the two MZI arms is recombined by the second 3-dB directional coupler (DC2) which terminates at the edge of the echelle slab waveguide, and launches light into the slab free-propagation region. The DC2 waveguides are close together at the slab boundary so that the two modes overlap as they propagate through the echelle to form Gaussian-like diffracted field profiles that closely match the modes of the output waveguides at the echelle focal line.

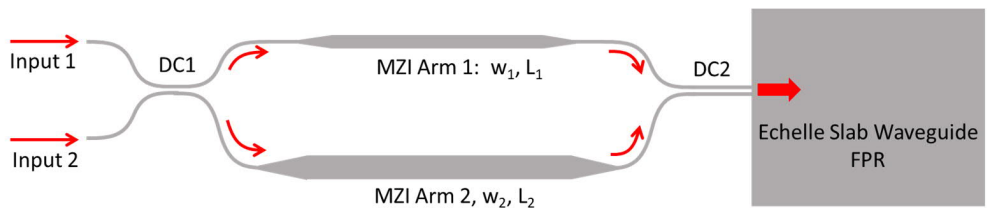

Fig. 2. Schematic (not to scale) of the temperature synchronized Mach-Zehnder interferometer. DC1 and DC2 are 3 dB directional couplers. The two output waveguides of DC2 launch light into the echelle slab waveguide free propagation region (FPR). The wide and narrow MZI waveguide sections with widths $w_{1,2}$ and lengths $L_{1,2}$ result in a temperature dependent phase difference $\Delta\phi$ between light arriving at the two DC2 waveguides. The remaining MZI arm waveguide sections are 500 nm wide and have equal total lengths in both branches so that they do not contribute to the MZI phase response.

The MZI waveguide arm widths for the athermal and super-thermal devices are different, with nominal widths of $w_1 = 500$ nm and $w_2 = 380$ nm for the devices fabricated and characterized in this work. The Si waveguide core has a much larger thermo-optic coefficient than the $SiO_2$ cladding. As a result, the narrower arm has a smaller waveguide thermo-optic coefficient $dN_{eff}/dT$ than the wide arm since the waveguide mode is less confined to the Si core. This thermo-optic imbalance causes the relative phase difference $\Delta\phi$ between light beams arriving at DC2, after propagating through the two arms of the MZI, to change with temperature. When $\Delta\phi$ changes, the ratio of optical power in the two DC2 output waveguides changes and the centroid of the combined output field envelope radiated into the free propagation slab section shifts laterally from one DC2 waveguide to the other, effectively changing the echelle incident beam angle $\theta$ at any point along the grating. Since the diffraction angles $\varphi_k$ in Equation 1 are fixed by the output waveguide positions along the echelle focal line, the shift in $\theta$ produces a corresponding temperature induced shift in the channel center wavelengths $\lambda_k$. Depending on the MZI orientation, the effective incident angle will either decrease of increase with temperature, leading to athermal or super-thermal behavior, as can be appreciated from Eq. (1). The third tunable MZI-echelle device is a simplified version of the layout shown in Fig. 2 that shares the identical DC2 structure as the athermal/super-thermal devices, but with balanced MZI waveguide arms of equal widths and lengths. A small metal strip heater is placed over one of the arms to tune the output phase difference $\Delta\phi$ of the MZI arms.

There are several design considerations that lead to the chosen MZI and DC dimensions in the athermal and super-thermal echelle devices. The theory and design procedure are described in Melati [13], and only the main results are summarized here. The DC2 waveguide widths and separation vary adiabatically to final widths $w_c = 255$ nm and separation $d_c = 950$ nm at the echelle slab FPR boundary, as shown in Fig. 3. These dimensions are a compromise between achieving the maximum possible lateral shift of the optical field distribution at the DC2 - FPR boundary (which gives the maximum channel wavelength shift), and generating an acceptable channel passband shape. The latter consideration requires that the two DC2 output waveguides are close enough together that the mode fields overlap to form a single Gaussian-like profile. The chosen DC2 waveguide separation allows the centroid of the launch mode spot to be shifted by up to 950 nm as the phase difference between the MZI arms varies from $\Delta\phi = -\pi/2$ to $\Delta\phi = \pi/2$. This displacement is sufficient to change the channel wavelengths by up to $\Delta\lambda = 1.6$ nm for the chosen Rowland circle radius, and thereby enables the MZI structure to correct the thermal channel wavelength shift over a temperature range of $\Delta T = 20$ K in the athermal device. This DC2 design gives a theoretical insertion loss of 0.4 dB and a smoothly varying single lobe channel passband shape with a 3 dB bandwidth of 250 GHz (2 nm), for MZI phase differences from $\Delta\phi = -\pi/2$ to $\Delta\phi = \pi/2$ [13]. As will be discussed in Section 3, outside this phase regime the light injected into the slab FPR region by the two DC2 waveguides interferes destructively resulting in a distorted passband shape and high insertion losses.

The MZI arm widths $w_1$ and $w_2$ and lengths $L_1$ and $L_2$ determine the temperature induced phase difference $\Delta\phi$ between light arriving at DC2 from two MZI arms, and hence the lateral displacement of the launched light distribution at the DC2 slab FPR junction. For the athermal echelle, the temperature dependent phase change must produce a spot shift that precisely cancels the spot shift at the echelle output waveguides caused by the slab waveguide thermo-optic effect. Recalling that the athermal temperature range of $\Delta T = 20$ K is determined a priori by the DC2 waveguide separation $d_c$ = 950 nm, the phase difference at the entrance to DC2 must change from $\Delta\phi = -\pi/2$ to $\Delta\phi = \pi/2$ as the temperature T varies over the operating range $T_0 - \Delta T/2 < T < T_0 + \Delta T/2$, where $T_0$ is the median operating temperature. Synchronization of the temperature induced phase difference to obtain an athermal device is achieved by choosing MZI arm lengths that satisfy the equation

$$L_1 \left.\frac{dN_{eff1}}{dT}\right|_{T_0} \cdot \Delta T - L_2 \left.\frac{dN_{eff2}}{dT}\right|_{T_0} \cdot \Delta T = \frac{\lambda_0}{2}. \tag{2}$$

where $dN_{eff1}/dT$ and $dN_{eff2}/dT$ are the waveguide effective thermo-optic coefficients of the two MZI arms [13], and $\lambda_0$ is the central operating wavelength for the demultiplexer.

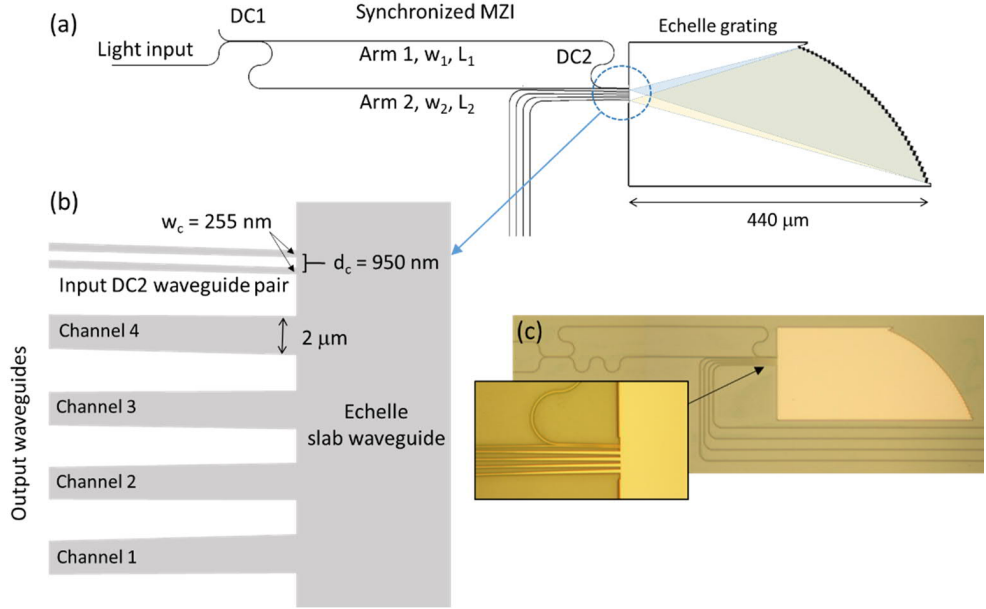

Fig. 3. (a) The chip layout of a combined MZI-echelle grating demultiplexer. (b) An expanded view showing details of the DC2 directional coupler and output channel waveguide arrangement at the focal line of the echelle grating slab waveguide FPR (c) Optical microscope images of a fabricated MZI-echelle device and the DC2 and output waveguide - FPR interface.

The second condition that determines the MZI waveguide widths and lengths is the requirement that $\Delta\phi$ should ideally be constant across the full operating wavelength range of the demultiplexer. This condition is necessary to ensure that the channel passband shapes are uniform across the four channel wavelengths, the wavelength correction by the MZI is the same across all channels, and also that the peak channel wavelengths remain close to the desired wavelength grid spacing across the operating wavelength band of the athermal and super-thermal demultiplexers. The condition $d\Delta\phi/d\lambda = 0$ for a wavelength independent MZI phase difference at $\lambda_0$ leads to the equation

$$n_{g1}(\lambda_0, T_0)\, L_1 = n_{g2}(\lambda_0, T_0)\, L_2, \tag{3}$$

where $n_{g1}$ and $n_{g2}$ are the MZI arm waveguide group indices. Eq. 3 is a useful guideline for choosing the MZI design parameters, and is often taken as the condition for a very wide MZI free spectral range (FSR). However, when group index dispersion is present, Eq. 3 is strictly satisfied at only one wavelength and does not guarantee a constant phase $\Delta\phi$ across all wavelength channels. Assessing the usable demultiplexer operating ranges and fabrication tolerances, as determined by the group index dispersion, will be addressed in Section 4.

Equations (2) and (3) can be simultaneously satisfied by a suitable combination of MZI waveguide arm lengths and waveguide widths that produce the necessary waveguide group index $n_g$ and effective waveguide thermo-optic coefficients [13]. The waveguide group indices $n_g$, effective indices $N_{eff}$, and effective thermo-optic coefficients were calculated using a two dimensional mode solver using thermo-optic coefficients $dn_{Si}/dT = 1.87 \times 10^{-4}$ [1] and $dn_{oxide}/dT = 1.5 \times 10^{-5}$ [22]. Based on these calculations, the MZI arm widths for the athermal and super-thermal devices were chosen to be $w_1 = 380$ nm and $w_2 = 500$ nm, with arm lengths of $L_1 = 2570$ μm and $L_2 = 2710$ μm respectively. The fabricated athermal and super-thermal devices both use the same MZI structure, except that the wide and narrow arms are interchanged with respect to the echelle axis (i.e. the horizontal axis in Fig. 2) in order to either cancel or amplify the intrinsic thermo-optic wavelength shift of the echelle grating. It is important to recognize that the MZI parameters obtained here by solving Eqs. 2 and 3 result in an unbalanced MZI with arms of unequal optical path length. In general the total cumulative phase difference $\Delta\phi_c$ between light arriving at the two arms of DC2 from the MZI is much larger than $2\pi$. In this paper the MZI phase difference is given as $\Delta\phi = \Delta\phi_c \mod 2\pi$, since this is sufficient to describe the athermal synchronization behavior, but in the discussion of fabrication tolerances in Section 4 it is important to consider the behavior of the cumulative phase $\Delta\phi_c$.

The design of the current tunable MZI-echelle device with the on-chip heater is much simpler than the athermal design. The overall layout is identical to the echelle-MZI device shown in Fig. 3, but with two MZI arms of equal lengths L= 600 μm and the waveguide widths both equal to w = 500 nm. A 185 μm long metal strip heating element is placed over one of the MZI arms so that the optical phase of light passing through that arm can be modulated using the thermo-optic effect by applying a small current to the heater. The light distribution at the DC2-echelle slab FPR boundary is thereby shifted from one DC2 waveguide to the other as in the case of the athermal device, and the demultiplexer channel wavelengths can be actively tuned over the full 1.6 nm of spectral tuning allowed by the DC coupler waveguide separation $d_c$. Unlike the athermal/super-thermal MZI design described previously, here the MZI is balanced and $\Delta\phi = 0$ over all temperatures and wavelengths when the micro-heater is off, assuming there are no intra-chip fabrication non-uniformities to unbalance the waveguides.

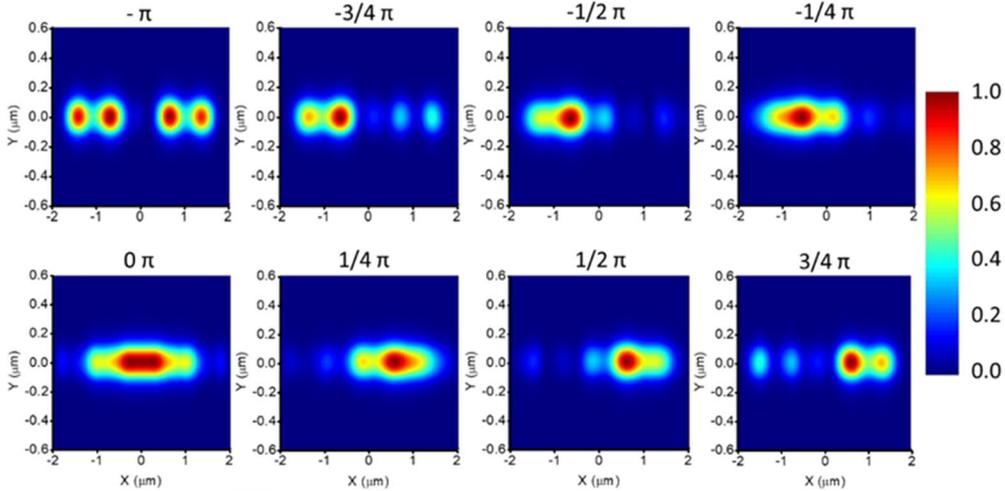

Fig. 4. The electric field mode distributions showing $|E|^2$ near the directional coupler-slab waveguide interface as the phase difference of the MZI arms varies from $\Delta\phi = -\pi$ to $\Delta\phi = \pi$.

## 3. Device Modelling

The performance of the echelle demultiplexers described above has been assessed by numerical simulations. Fig. 4 shows the TE mode electric field evolution at the DC2-echelle slab FPR interface as the phase difference after propagation through the two MZI arms varies from $\Delta\phi = -\pi$ to $\Delta\phi = \pi$. These field profiles are obtained by using a finite difference based eigen-mode expansion (EME) propagation method [23]. The displayed $|E|^2$ profiles are taken at a cross-sectional plane located at 2 µm from the DC2 waveguide termination inside the echelle slab waveguide FPR. The centroid of the field distribution shifts from left to right as the phase difference varies from $\Delta\phi = -\pi/2$ to $\Delta\phi = \pi/2$. These phase boundaries correspond to temperatures of T = 290 K and T = 310 K in the present MZI design. As noted previously, phases less than $\Delta\phi = -\pi/2$ and greater than $\Delta\phi = \pi/2$ are not useful. As $\Delta\phi$ approaches $\pm\pi$ the light in the two DC2 waveguides grows increasingly out of phase, and interferes destructively between the two DC2 waveguide outputs. The overlap between the diffracted spot with the output waveguides deteriorates, resulting in a notched passband and high insertion losses.

The demultiplexer channel passband shapes and wavelengths were calculated by using an in-house echelle grating design software [13]. Fig. 5 shows the predicted channel passbands at temperatures (phases) of T = 290 K ($\Delta\phi = -\pi/2$), 300 K ($\Delta\phi = 0$) and 310 K ($\Delta\phi = \pi/2$) for the simple echelle device and for the athermal and super-thermal echelle demultiplexers with the co-integrated MZI. At 300 K, the four channel wavelengths are $\lambda_1 = 1540.5$ nm, $\lambda_2 = 1546.9$ nm, $\lambda_3 = 1553.3$ nm, and $\lambda_4 = 1559.8$ nm. The passband simulation demonstrates that all the field distributions in Fig. 4 between $\Delta\phi = -\pi/2$ to $\Delta\phi = \pi/2$ produce almost identical channel passband shapes after propagation through the echelle demultiplexer. The predicted insertion loss penalty in the athermal and super-thermal MZI-echelle devices, relative to the simple echelle, is better than -2 dB for all channels, and channel cross-talk is less than -30 dB. The temperature induced channel wavelength shift of the basic echelle device in Fig. 4(a) is 80 pm/K or 1.6 nm across the $\Delta T = 20$ K range shown here. In the athermal device the temperature induced channel shift is almost completely cancelled with a residual change of less than 50 pm across the 20 K range. Finally, the super-thermal device shows that the channel shift is enhanced to 160 pm/K, or 3.2 nm across the 20 K range, and also retains excellent passband fidelity. This shift is exactly double the shift of the basic echelle in Fig. 4(a) because the super-thermal device uses an identical MZI as the athermal device but with the MZI arms interchanged in order to amplify rather than cancel the intrinsic temperature induced channel wavelength shift. The super-thermal device could be designed to have a larger wavelength tuning rate by choosing MZI arms with a larger temperature induced phase shift $d\Delta\phi/dT$, but in such a device the maximum wavelength shift is still limited to 3.2 nm by the choice of DC2 waveguide separation $d_c$ and the echelle Rowland radius.

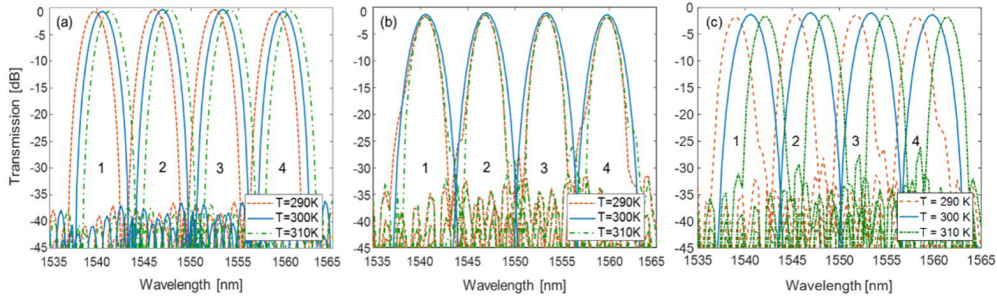

Fig. 5. The calculated demultiplexer channel passbands at T = 290, 300, and 310 K, for (a) the basic echelle as in Fig. 1, (b) the co-integrated MZI-echelle in athermal configuration as in Fig. 3, and (c) the MZI-echelle in the super-thermal configuration. In (b) and (c), these three temperatures correspond to MZI arm phases of $\Delta\phi = -\pi/2$, $\Delta\phi = 0$, and $\phi = \pi/2$, respectively.

For simplicity in establishing the relationship between input field profiles of the DC2 output shown in Fig. 4 and output passband shape, the MZI-echelle simulation model used to generate the plots in Fig. 5 does not include group index dispersion. The latter will cause a corresponding wavelength dispersion in MZI phase difference Δϕ at any given temperature, but does not change the transformation between DC2 output field profiles of Fig. 4 and the output passband shape for any given Δϕ. The effect of dispersion on real device performance is discussed in the following section.

## 4. Tolerance Analysis

This section examines the effects of dimensional variations and wavelength dispersion on the performance of the MZI-echelle devices. The relationship between fabrication imperfections and echelle and AWG demultiplexer cross-talk and insertion loss has already been extensively examined in the demultiplexer literature [e.g. 17,24,25]. The focus here is on the MZI structure and the effect of waveguide effective index dispersion on dimensional tolerances and device wavelength and temperature operating ranges. Dimensional tolerances are set by requiring that a fabricated device meets specifications for channel wavelengths and insertion loss over a desired temperature range $T_0 - \Delta T/2 < T < T_0 + \Delta T/2$ centered at the median operating temperature $T_0$. This is equivalent to requiring that at the median temperature $T_0$, the phase difference Δϕ remains much less than π across the full demultiplexer wavelength band. Dimensional tolerances and the role of wavelength dispersion are obtained by directly evaluating the wavelength dependent MZI output phase difference $\Delta\phi(\lambda,T_0)$ across the demultiplexer wavelength operating band for different MZI arm waveguide widths, using exact mode effective indices $N_{eff}$ calculated at each wavelength. In this way, effective index and group index dispersion are naturally built into the analysis with no a priori approximations.

Single lobed channel passbands are only obtained for phases in the range $-\pi/2 < \Delta\phi < \pi/2$, since outside this range the beam profile launched into the echelle free propagation region by DC2 displays multiple peaks and notches as illustrated in Fig. 4. The calculated echelle channel passbands shown in Figs. 5(b) and 5(c) span this full range for the experimental devices in this work, since by design the ΔT=20 K temperature change corresponds to exactly this phase interval. It is useful to consider $\psi_{\Delta T}$, the total MZI phase excursion that would be expected over the design temperature range, $T_0 - \Delta T/2 < T < T_0 + \Delta T/2$. The MZI output phase difference at $T_0$ must then satisfy the relation

$$-\left|\frac{\pi}{2} - \frac{\psi_{\Delta T}}{2}\right| < \Delta\phi(\lambda, T_0) < \left|\frac{\pi}{2} - \frac{\psi_{\Delta T}}{2}\right|, \qquad (4)$$

across the demultiplexer wavelength band in order to ensure acceptable passband shape and insertion loss over the full ΔT temperature span, uniform temperature correction for each channel, and acceptable deviation of wavelength channels from the design grid spacing. In the ideal situation that $\Delta\phi(\lambda,T_0) = 0$ at $T_0$ across entire wavelength band, Eq. 4 predicts that the maximum temperature operating range is given by the condition $\psi_{\Delta T} = \pi$. The athermal or super-thermal temperature operating range is then limited only by the DC2 waveguide separation $d_c = 950$ nm. This is the basis of the design presented in the previous sections and in [13]. On the other hand, in the presence of group index dispersion Eq. 3 is only satisfied at one wavelength so the phase difference $\Delta\phi(\lambda,T_0)$ may vary significantly with wavelength, and either the temperature operating range or the wavelength operating band must be reduced so that Eq. 4 is satisfied.

Group index dispersion and hence dispersion in $\Delta\phi(\lambda,T_0)$ vary with Si waveguide width and thickness. Fig. 6(a) shows the $\Delta\phi(\lambda,T_0)$ dispersion across the C-band for several MZI waveguide arm width combinations corresponding to fabricated waveguides that are slightly wider or narrower than the nominal MZI design given above (i.e. $w_1 = 500$ nm, $L_1 = 2570$ μm,

$w_2 = 380$, $L_2 = 2710$ μm). The examples cover a 30 nm width range that encompasses the ±10 nm variation commonly encountered in Si waveguide fabrication. Fig. 6(b) shows $\Delta\phi$ dispersion curves for 220 nm, 210 and 230 nm Si waveguide thicknesses representative of possible 5% Si thickness variations around the 220 nm design value. While in-wafer thickness uniformity is well within ±1 nm using modern SOI manufacturing processes, a recent study [26] has found that the difference in Si thickness from the nominal 220 nm and from wafer to wafer varied randomly by up to 9 nm over a small sample of commercial SOI wafers. The phase curves in Fig. 6 were calculated using waveguide effective index values obtained using a finite difference mode solver [23]. The bulk Si and $SiO_2$ refractive index values (at T = 300 K) used in the mode solver were interpolated from data in Palik [27]. For illustrative purposes, the shaded regions in Fig. 6 outline a notional tolerance window corresponding to a $\Delta\phi$ variation of $\pm\pi/4$, and a wavelength band of 25 nm to include all four channels of the present design. If a temperature compensated operating range of at least $\Delta T = 10$ K is required, the $\Delta\phi(\lambda, T_0)$ curves must fall entirely within this $\pi/2$ wide tolerance window over the full wavelength band. To highlight the effect of phase dispersion in this discussion, rather than absolute phase which will be discussed below, each curve in Fig. 6 has been offset vertically by a constant so that they all pass through the center of the tolerance window. This is equivalent to artificially setting the median temperature $T_0$ to be the same for each example.

The $\Delta\phi$ curve in Fig. 6(a) for a MZI with $w_1 = 490$ nm and $w_2 = 370$ nm varies by $0.15\pi$ across the 1535 to 1560 nm operating band. Eq. 4 implies that the full wavelength band can therefore only be used over a temperature range such that $\psi_{\Delta T} < 0.85\pi$. The usable athermal temperature range of the MZI-echelle demultiplexer is then limited to 85% of the maximum possible 20 K range set by the DC2 waveguide separation, or $\Delta T \sim 17$ K. This example gives the best performance that can be expected from the current design, since the point of zero dispersion is centered on the operating band and dispersion across the band is the minimum possible. By a similar argument, the 500 nm / 380 nm MZI will be limited to a $\Delta T = 14$ K given the $\Delta\phi$ variation of $0.3\pi$ in Fig. 6(a). At the other extreme, the $\Delta\phi$ curve in Fig. 6(b) for 210 nm thick silicon waveguide varies by almost $\pi$ across the operating band. Eq. 4 is only satisfied across the full wavelength band if $\psi_{\Delta T}$ and therefore $\Delta T$ are close to zero. In this example even a small temperature change will cause some channel passbands to become distorted and have high insertion loss. Furthermore, even at temperature $T_0$ the large phase variation across the band will cause some channel wavelengths to be displaced by as much as ±0.8 nm from the nominal wavelength grid.

Inspection of the curves in Fig. 6(a) suggests that acceptable athermal or super-thermal demultiplexer performance over a 10 K range as delineated by the $\pm\pi/4$ tolerance window can be achieved for MZI waveguide widths within ±10 nm of the nominal 500 nm/380 nm design. Although the 490nm/370 nm MZI arm width combination displays the smallest phase dispersion in this analysis, the dispersion is much larger for the narrower 480 nm/ 360 nm waveguide arms. Therefore the 500nm/380 nm MZI may be the better choice given typical fabrication variations. Similarly the phase dispersion curves in Fig. 6(b) show that the phase dispersion is also sensitive to waveguide thickness, which should be within ±5 nm of the nominal 220 nm specification to fall within the example tolerance window.

The second fabrication tolerance consideration is to obtain a fabricated demultiplexer for a desired median operating temperature $T_0$. This is equivalent to requiring that $\Delta\phi$ should be near zero at temperature $T_0$ and at the center wavelength $\lambda_0 \sim 1550$ nm. From Eq. 4, this condition should give the largest possible temperature operating range before passband distortion occurs. Unfortunately it is impossible to predict the exact operating $T_0$ (i.e. the temperature at which $\Delta\phi = 0$) after fabrication, for the current athermal and super-thermal designs. As noted previously, the MZI geometry obtained by solving Eqs. 2 and 3 will not have balanced optical path lengths in the two arms. For example in the athermal design presented in this paper, the optical path length difference between the two arms is equivalent to several hundred wavelengths. As a result, the cumulative phase difference $\Delta\phi_c$ is extremely sensitive to small changes in effective index. For the nominal 500 nm/380 nm MZI design, the accumulated phase $\Delta\phi_c$ changes by approximately $4.5\pi$ per nm of waveguide width change. Small waveguide thickness variations will also result in similar large phase variations. Given that fabricated Si waveguide widths and thicknesses can deviate from the design value by several nanometers with current fabrication processes, the value of $\Delta\phi$ will fall anywhere between $-\pi$ and $\pi$ for any specific temperature and wavelength. Fortunately the phase $\Delta\phi$ is periodic in temperature, with a period of $2\Delta T$. Furthermore, $\Delta\phi$ can be shifted by $\pi$ simply by changing which of the two possible DC1 directional coupler waveguides is used as an input. An effective $T_0$ for any given fabricated device can always be found within a temperature interval less than $\Delta T$ from the nominal design value of $T_0$. As a result, a device with specific median operating temperature $T_0$ can usually be obtained by post-fabrication selection from a reasonably small sample of fabricated devices. While post-fabrication selection is acceptable for research purposes, the low yield is obviously not desirable for commercial manufacturing. A more reproducible design may require new waveguide structures such that Eq. 3 can be satisfied while minimizing the optical path length difference of two MZI arms. Possible future approaches could be to use waveguide structures with a much larger thermo-optic coefficient difference than possible with the simple photonic wires used here [28], or to use dispersion engineered waveguides [29,30] where both the first and second order effective index dispersion and second order dispersion are suppressed. Alternatively, some form of post-fabrication or in-situ tuning of the MZI arm may be possible to bring the MZI phase difference into specification at the desired median operating temperature.

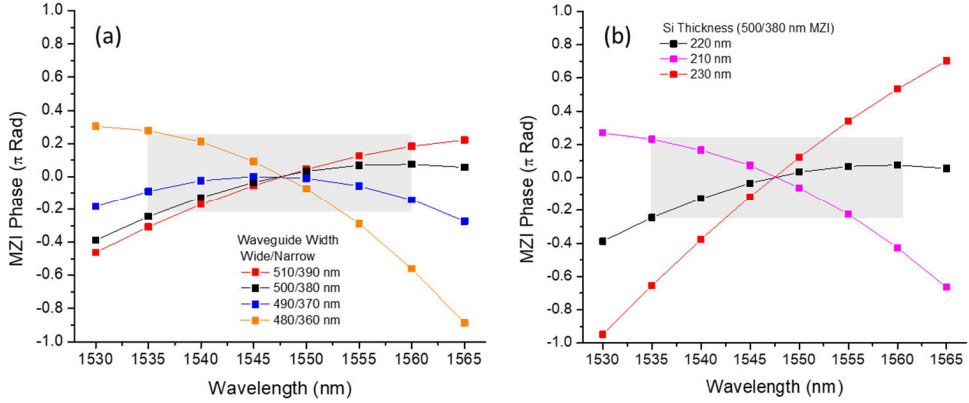

Fig. 6(a). The calculated MZI phase difference $\Delta\phi$ for MZI arm waveguide widths varying from +10 nm to -20 nm from the nominal $w_1 = 500$ nm, $w_2 = 380$ nm design, at a fixed 220 nm Si waveguide thickness. (b) The calculated $\Delta\phi$ for Si waveguide thickness varying from +10 nm to -10 nm from the nominal 220 nm thickness, for fixed $w_1 = 500$ and $w_2 = 380$ nm arm widths. The shaded region represents a $\pm\pi/4$ tolerance window.

Fabrication tolerance considerations for the actively tunable MZI-echelle device with an on-chip heater are considerably simpler. Since the MZI arms are identical, any variation in fabricated waveguide width and thickness from the nominal design will not cause significant dispersion in $\Delta\phi$ provided that the dimensional error is nearly the same for both arms. This is usually the case as long as the features in question are in the same area of the chip, with no nearby features that may perturb local lithography exposure and etching rates. Hence $\Delta\phi$ will be nearly constant across the demultiplexer wavelength band. Nevertheless even very small differences in dimensions of the two arms may cause some deviation from the ideal $\Delta\phi = 0$ value. However since this device is tunable, any intrinsic phase offset is easily corrected.

## 5. Experiment

The athermal, super-thermal, and on-chip tunable MZI-echelle devices were fabricated on an SOI chip with a 220 nm thick Si waveguide layer, a 2 μm thick buried oxide layer and a 2.2 μm SiO$_2$ upper cladding. All nominal device dimensions are as given in Section 2. Patterning was done by e-beam lithography and waveguides were formed by reactive ion etching. The heater element for the MZI–echelle with on-chip tuning is a 185 μm long and 5 μm wide TiW alloy metal strip centered over one MZI waveguide arm, and is approximately 200 nm thick. The measured resistance of the heater, including the contact pad and leads, was 180 ohms. All devices were tested using a fiber coupled tunable laser with 1 mW power. TE polarized light was coupled on and off the chip through polarization maintaining fibers, and measured with a fiber coupled photodiode. On-chip Si subwavelength grating edge couplers [31] were used at the chip facet for efficient waveguide to fiber mode matching. During testing the chips were mounted on a temperature controlled stage with a temperature accuracy of 0.1 °C.

The basic echelle configuration as in Fig. 1 was characterized first, to assess the base line performance of the stand-alone echelle grating. The measured channel spectra are shown in Fig. 6(a) for temperatures of 20 °C and 50 °C. The measured total insertion losses through t 9 mm long 500 nm wide straight reference waveguides on the same chip were -10 dB ( ±1 dB). This includes -2 dB of fiber-to-fiber system loss, the input and output fiber-to-chip coupling losses, and on-chip waveguide losses of -3 dB/cm. In comparison with the reference waveguides, the basic echelle device introduces less than 2 dB of additional insertion loss at the channel center wavelengths. The temperature induced shift of channel wavelengths between 20 °C to 50 °C was 2.5 nm, which corresponds to an intrinsic channel temperature dependence of 83 pm/°C for the basic echelle design.

The athermal MZI-echelle demultiplexer channel spectra are shown in Fig. 7(b) for several temperatures between 20 °C and 45 °C. In comparison with the basic echelle, the MZI structure adds -2 dB to the insertion loss. Fig. 8 shows the temperature induced shift of all four athermal device channels. The channel wavelengths are taken as the midpoint between the two -3 dB power drop-off wavelengths on each side of for the measured channel passbands. For the athermal device the temperature induced shifts for channels 2 and 3 are less than 200 pm between 20 °C and 40 °C, compared with the 1.6 nm thermo-optic shift for the basic echelle demultiplexer. For these two channels the insertion loss variation with temperature is less than 1 dB. The outlying channels 1 and 4 show a larger temperature shifts of up to 500 pm and an insertion loss variation with temperature of up to 3dB. The bandwidth of athermal operation over a full 20 °C range with minimal passband variation over temperature is therefore limited an approximately 15 nm wavelength window in the middle of the C-band. If a smaller athermal temperature range is acceptable for an application, then for all four channels the insertion loss varies by less than 1.5 dB over the more restricted 15 °C temperature span between 25 °C to 40 °C. The insertion loss and pass band shape distortions are more significant for the first and last channels due to the presence of MZI phase dispersion, as discussed in Section 4. The passband distortion and additional insertion loss are particularly evident for channel 4 at 20 °C and channel 1 at 45 °C. A separate MZI test structure, identical to the MZI used in this athermal MZI-echelle, has a measured free spectral range of 50 nm. Therefore the phase difference $\Delta\phi$ between MZI arms varies by $2\pi$ over 50 nm and hence by about $\pi$ across the 25 nm four channel range of this device. As expected from Eq. 4, only the channels in a much narrower wavelength band remain relatively undistorted over the full 20 °C temperature range.

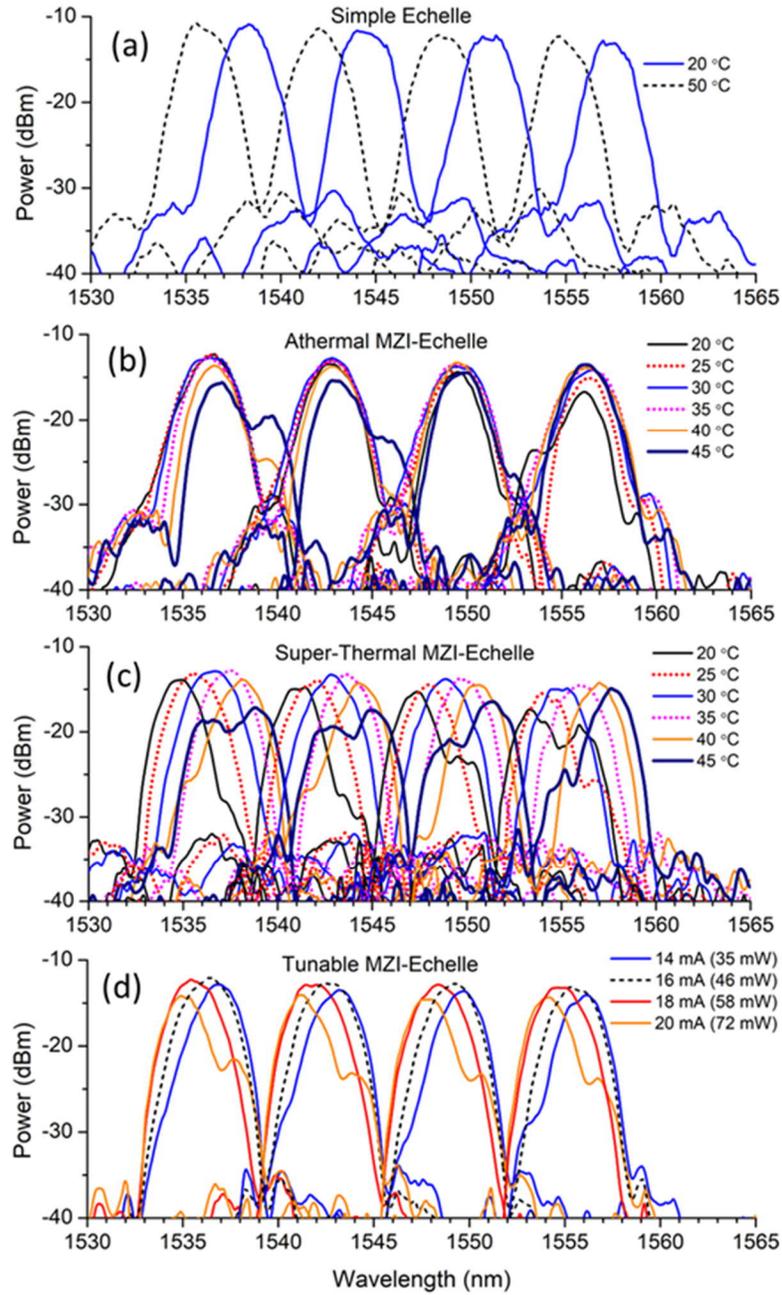

Fig. 7. Measured echelle grating channel spectra at various temperatures for (a) a single input waveguide demultiplexer as in Fig. 1 and no MZI launch structure, (b) an athermal MZI-echelle with the temperature synchronized MZI launch structure, (c) a super-thermal MZI-echelle, and (d) the tunable MZI-echelle with on-chip heater phase control. The legend shows applied on-chip heater current and heater power at each setting.

The measured channel passbands for a super-thermal device are shown in Fig. 7(c), and the corresponding temperature induce channel wavelength shifts are again shown in Fig. 8. This device is identical to the athermal device of Fig. 7(b), except that the wide and narrow MZI arms are interchanged relative to the two DC2 output waveguides (i.e. in Fig 2, the upper and lower MZI arms are interchanged). The channel passband spectra show clear super-thermal behavior with an enhanced temperature induced channel tuning of approximately 170 pm/°C, as expected, since the athermal MZI parameters were chosen to exactly compensate for the 83 pm/°C channel shift of the basic echelle. As in the case of the athermal device, the usable wavelength span for a full 20 °C temperature tuning range is limited by dispersion. Channels 1 through 3 show super-thermal shifts with relatively undistorted passbands over a full 20 °C temperature range, with a peak insertion loss change of less than 1.5 dB from 20 °C and 40 °C. Over a smaller 15 °C range between 25 °C and 40 °C, all four channels display insertion loss variation with temperature of less than 1.5 dB.

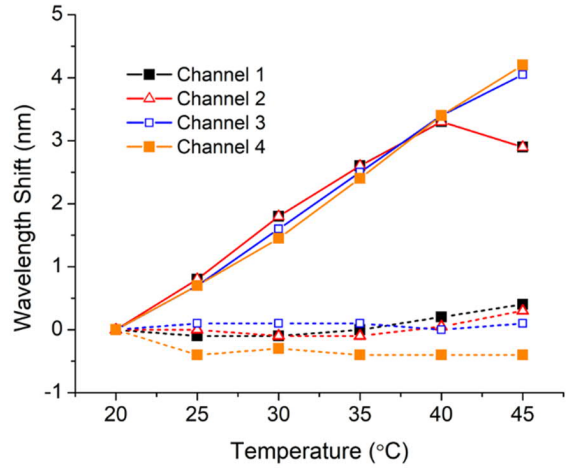

Fig. 8. Wavelength shift of the channel passband center wavelengths with temperature for the athermal (dashed lines) and super-thermal (solid line) devices. All shifts are shown relative to the channel wavelengths measure at 20 °C.

In the final example, Fig. 7(d) shows the spectra for the MZI –echelle device with on-chip tuning. The underlying stage temperature was held at 20 °C for these measurements. Heater currents from 0 to 100 mA were applied using a computer controlled current source. Fig. 7(d) shows the channel spectra for applied heater currents ranging from I = 14 mA to 20 mA, corresponding to heater powers from 35 mW to 72 mW. Note that the 14 mA baseline current was required to bring the MZI arms into the correct phase relationship, due to a small residual MZI optical path imbalance. The on-chip heater shifts channel wavelengths at a rate of 38 pm/mW. The device channels of Fig 7(d) can be tuned over $\Delta\lambda$ = 1.6 nm using less than 80 mW heater power consumption. The measured channel shifts are the same for all channels, to within the 0.1 nm measurement uncertainty. There is negligible distortion of the channel passband shape and an insertion loss variation of less than 1 dB across all four channels. Given that the intrinsic echelle temperature dependence is 83 pm/°C, active tuning can stabilize the demultiplexer channel positions over the equivalent a 20 °C chip temperature change. For comparison, maintaining the temperature of a chip and carrier at a differential of several degrees °C from ambient using a thermoelectric cooler typically consumes a few Watts [32]. The maximum tuning range is still fixed by the DC2 waveguide separation, and currents outside the

14 to 20 mA range in Fig 7(d) cause the MZI phase difference to fall outside the allowed $\pm\pi/2$ range, which results in passband distortion. This is illustrated by the spectra for I = 20 mA in Fig. 7(d), which corresponds to the $\Delta\phi = 3/4\pi$ input field profile distribution in Fig. 3. The uniformity of the channel spectra across the C-band in Fig. 7(d) confirms that the wavelength operating range of this tunable echelle device is not limited by the dispersion of $\Delta\phi$ since the MZI arms have the same widths and lengths. Application of a heater current does shift the MZI from the perfectly balanced condition but the change is too small to cause obvious channel pass-band shape variation across the four wavelength channels.

## 6. Summary

Temperature dependent MZI launch structures have been demonstrated in three configurations to control the temperature dependent response of echelle grating demultiplexer output channel wavelengths. The MZI was used to create a true athermal device, and also super-thermal and on-chip tunable silicon photonic echelle grating demultiplexers that have power efficient tuning mechanisms to actively lock the wavelength channels to the grid. The previously published theoretical analysis [13] has also been extended to incorporate the effect of group index dispersion, which is shown to be essential in understanding device performance and fabrication tolerances. Finally, experimental results are presented for fabricated athermal, super-thermal and on-chip tunable silicon photonic MZI-echelle grating devices. The athermal device has a channel temperature dependence of less than 10 pm/°C over a restricted operating range, as compared to the 83 pm/°C shift observed for the corresponding simple stand-alone echelle. Athermal operation of the fabricated device without passband degradation over the full 20 °C is only achieved over a limited 15 nm wavelength range encompassing two channels, or for all four channels over a smaller 15 °C wide temperature span. These operating range limitations result from wavelength dispersion of the MZI phase difference $\Delta\phi$. Phase dispersion is never completely absent for these Si waveguides but can be minimized by choosing the MZI parameters to give $d\Delta\phi/d\lambda=0$ at the center operating wavelength. However, achieving this in practice is difficult because dispersion changes rapidly with waveguide width and thickness deviations from nominal design values after fabrication. The tunable super-thermal design suffers from the same sensitivity to fabrication variations. On the other hand the tunable MZI –echelle with on chip heater avoids this problem since the MZI arms are balanced and there is no dispersion of the output phase. This MZI launch structure can shift the channel wavelengths up to 1.6 nm under active electrical control, a range sufficient to compensate chip temperature variations of up to 20 °C. Since the thermo-optic modulator only heats one MZI waveguide arm, the total power required to actively tune the channel wavelengths is significantly less than that would be required to tune a simple echelle by changing the temperature of a full chip and underlying mount. As a result less heat will be dissipated into the chip and local environment, reducing the cooling requirements for the entire system. Finally, the on-chip MZI tuning scheme allows several demultiplexers on the same chip to be independently tuned. In addition to thermal compensation, this capability can allow deviations in device wavelength channels to be independently corrected.

All three MZI-echelle configurations demonstrated in this paper are potential solutions for stabilizing wavelength demultiplexers in a fluctuating thermal environment. The fully athermal demultiplexer design as originally proposed [13] would be the most attractive solution, since it completely removes the need for active temperature control. However, the fabrication tolerances needed to guarantee a wide operating wavelength and wide temperature range are challenging given current fabrication technology, particularly because of the strong group index dispersion in Si photonic wire waveguides. On the other hand, the MZI-echelle with on–chip tuning allows for low power wavelength channel tuning to correct for environmental temperature fluctuations, and displays good immunity to fabrication variations. In situations were active on-chip tuning is an option, the tunable MZI-echelle is an attractive and practical choice.

**Disclosures.** The authors declare no conflicts of interest.

**Data availability.** Data underlying the results presented in this paper are not publicly available at this time but may be obtained from the authors upon reasonable request.